# AGING AS A CONSEQUENCE OF MISREPAIR
## -- A NOVEL THEORY OF AGING

Jicun Wang[1], Thomas M. Michelitsch[2*], Arne Wunderlin[3] and Ravi Mahadeva[1]

[1]Department of Medicine, University of Cambridge, Addenbrooke's Hospital, Cambridge, United Kingdom
[2]Institut Jean le Rond d'Alembert CNRS UMR 7190, Université Pierre et Marie Curie (Paris 6), 4 Place Jussieu, 75252 Paris cedex 05, France
[3]1. Institut für Theoretische Physik, Pfaffenwaldring 57/4, D-70550 Stuttgart, Germany
*Send correspondence to: Dr. Thomas M. Michelitsch, Email: michel@lmm.jussieu.fr

27 Mars 2011

**ABSTRACT**

It is now increasingly realized that the underlying mechanism which governs aging (ageing) is a complex interplay of genetic regulation and damage-accumulation. "Aging as a result of accumulation of 'faults' on cellular and molecular levels", has been proposed in the damage (fault)-accumulation theory. However, this theory fails to explain some aging phenotypes such as fibrosis and premature aging, since terms such as 'damage' and 'fault' are not specified. Therefore we introduce some crucial modifications of this theory and arrive at a novel theory: *aging of the body is the result of accumulation of Misrepair of tissue*. It emphasizes: **a)** it is Misrepair, not the original damage, that accumulates and leads to aging; and **b)** aging can occur at different levels, however aging of the body takes place necessarily on the tissue level, but not requiring the aging of cells/molecules. The Misrepair-accumulation theory introduced in the present paper unifies the understanding of the roles of environmental damage, repair, gene regulation, and multicellular structure in the aging process. This theory gives explanations for the aging phenotypes, premature aging, the difference of longevity in different species, and it is consistent with the physical view on complex systems.

**INTRODUCTION**

Many theories have been proposed to answer the questions of why and how we age. These theories fall into two main groups: genetic regulation and damage-accumulation. The **gene-controlling theories** emphasize the genetic controlling on aging (ageing). Many genes have been found to be related to aging, and some have proven to have lifespan-extending effects on animal models by gene knock-out (or -in), however their associations with aging are debated and far from confirmed. The questions are, if there are such kinds of genes: **a)** how their functions are related to the variety and inhomogeneity of the phenotypes of aging? **b)** why a living being has such kinds of genes to make itself going to die? and **c)** since when (which age) they exert functions badly? Among these theories, the *Mutation-accumulation Theory* assumes that the weak selection by nature in late age allows a wide range of gene mutations to accumulate, with deleterious effects (Medawar, 1952 and Martin, 1996): one interesting point which remains unanswered is how cells can survive and function well when genes mutate with deleterious effects. The *Developmental Theory* suggests that aging and development are coupled and regulated by the same mechanisms (Zwaan, 2003 and De Magalhaes, 2005). It is interesting and reasonable; however this theory does not explain how the development process influences the aging phenotypes. The **Damage-accumulation theories** imply that it is the extrinsic and intrinsic damage (or fault) -accumulation on cellular or molecular level that leads to aging: The causes can be free radicals, by-products, molecular cross-linking and so on (Kirkwood, 2000 and Holliday, 2004). However, most of these theories actually mainly pointed out the phenotypes or causes but not the *underlying common mechanism* of aging.

It is now increasingly realised that genetic regulation and damage-accumulation actually interplay to cause aging, and it is predicted that: *"aging is a result of accumulation of 'faults' at cellular and molecular level because of the limitation of maintenance and repair; the underlying driving force is damage. The genetic* *control of longevity comes through the regulation of the essential maintenance and repair processes that slow the build–up of faults"* (Kirkwood, 2006). However this recent damage (fault)-accumulation theory does not clarify the concepts of 'damage' and 'fault', and is weak in explaining some aging phenomena, such as fibrosis and premature aging. Therefore we suggest here some crucial modifications of this theory and raise a novel theory of aging.

## A NOVEL THEORY OF AGING

### Limitations of the current damage (fault)-accumulation theory

In the above mentioned damage (fault)-accumulation theory (Kirkwood, 2006), the terms of 'damage' and 'fault' are not clearly defined. The 'damage' can mean the primary damage before reparation and/or the incomplete repair following damage. The damage can be extrinsic or intrinsic. Incomplete repair may also include partial repair and incorrect repair. However it is necessary to distinguish between them. For example, the burnt skin develops a scar. The burnt skin is the damage, and the scar is the incorrect repair. The 'fault' seems to mean the original damage which is either partially repaired or not at all repaired, by mistake due to the limitation of repair system. However, if the damage is left unrepaired, it will develop a life-threatening condition, such as bleeding, infection and organ failure, by destroying the integrity of structure, no matter whether the damage is on the molecular, cellular, or tissue level. Therefore, the original damage cannot remain unrepaired, and hence the concept of accumulation of 'faults' is misleading. Reparation of any damage, and reconstruction of the structure, is essential to maintain the integrity of structure and the basic function for survival: even if sometimes it is imperfect and lead to a reduction of function. In most cases, reparation is complete, so that the structure can be restored. However, incorrect repair can and even must occur in cases of serious or frequent injuries, where quick reparation is more important for the immediate survival than perfect reparation. Typical examples include the Misrepair of DNA after double-strand breaks of DNA (Kevin, 2004) and the scar formation. The term 'Misrepair' appears applicable in a wider sense, not only to DNA, but also to other types of incorrect repair due to a similar mechanism.

### The concept of Misrepair

We define the concept of Misrepair as *the incorrect reconstruction of a structure*, *after* repairing the original damage, which can lead to a change of structure (of molecule/cell/tissue) and a reduction of function. The change of structure can be the change of the kinds or amounts of functional components, and their locations or spatial relationships. The comparison between damage and Misrepair is shown in Table 1. We define the term "damage" as the structure change before any reparation has taken place, whereas "Misrepair" is the change of structure as a result of incomplete or defective reparation. Distinguishing these two concepts is necessary to understand the aging mechanism, in which damage is the cause and Misrepair is the effect. Misrepair is unavoidable when serious or frequent damage occurs. For example, the non-stopping mechanical movements of heart wall, arterial wall and skin often cause the break of muscle fibres, elastic fibres and other extracellular matrixes (ECMs). Misrepairs on different structures appear in different manners, such as change of DNA sequence, rearrangement of cytoskeleton (Garcia, 2002), Mallory body formation, rearrangement of ECMs (Cook, 2000), cellular reorganization (Eliasieh, 2007), and tissue fibrosis. Most of the degeneration changes, such as hyaline degeneration (Mallory body), mucoid degeneration, pigment particle, and calcium homeostasis are in fact the products of Misrepair.

**Table 1. Comparison of damage and Misrepair**

| Damage | Misrepair |
| --- | --- |
| Physical and chemical injury on molecule/cell/tissue | Imperfect reconstruction of structure of molecule/cell/tissue |
| Defect before reparation | Defect after reparation |
| Removable | Permanent |
| e.g. Injury or death of molecule/cell/tissue | e.g. Change of the kinds or amounts of functional components and their locations or spatial relationships |

Actually, apart from the Misrepair of DNA, *Defective Repair, Adaptive Repair, and Molecular and Cellular Reorganisation*, which have similar meaning to Misrepair, have been recognised in cells and molecules (Bansal, 2003, Cenacchi, 2005, and Eliasieh, 2007). A tissue Misrepair hypothesis was also raised to explain the radiation carcinogenesis (Kondo, 1991). However the importance of Misrepair mechanism was not sufficiently emphasized and its association with aging has not yet been studied. Here we discuss mainly the importance of Misrepair in aging.

Misrepair occurs no matter how capable the gene-regulated reparation and -maintenance system is, since reparation – many steps of biochemical reactions – needs time. Misrepair is a result of the active reparation processes beneficial for survival, but not by mistake due to the limitation of repair system. Therefore the maintenance and repair system does NOT SLOW but even PROMOTES the occurrence of Misrepair if necessary, although a more capable repairing system means lower rate of occurrence of Misrepair. Hence the appearance of Misrepair gives more chance for survival of individual, by which the individual can live at least up to the reproduction age, which is important for the survival of species. Therefore the Misrepair mechanism was selected by nature due to its evolutionary advantage.

**Aging is the result of accumulation of Misrepair**

A misrepaired structure cannot be 'repaired' any more (restored to the pre-damage state) and persists as a 'scar'. So it is the Misrepair and not the original damage that accumulates. Therefore the term of *Misrepair-accumulation* is more appropriate than that of damage-accumulation (Figure 1). Damage drives the aging process by triggering Misrepair. The accumulation of Misrepair leads to the gradual de-organisation of structure and the gradual reduction of function. Liver cirrhosis, for instance, is a result of accumulation of mis-reconstruction of tissue due to the death of hepatocytes in chronic hepatises. Aging is a result of reparation, but not a result without reparation. Aging is necessary for body immediate survival and species survival. This is why evolution favours aging mechanism. Thus Misrepair might represent the mechanism by which *organisms are not programmed to die but to survive* (as long as possible) (Kirkwood, 2005), and aging is just the price to be paid.

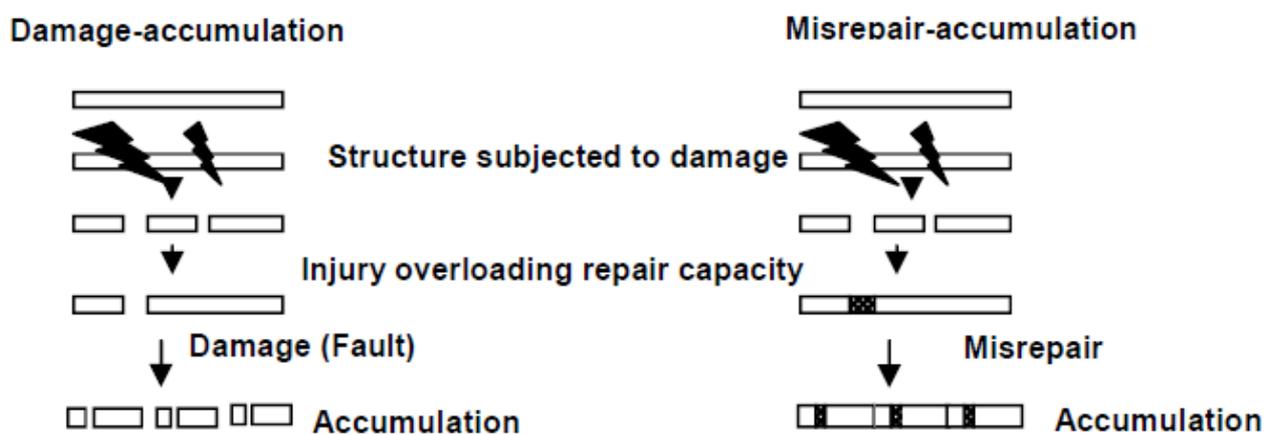

**Figure 1. Comparison of Misrepair-accumulation and damage-accumulation**

**Damage-accumulation:** due to the limitation of maintenance and repair, some damage remains partially repaired or unrepaired by mistake. This remained damage (fault) will accumulate and lead to aging. **Misrepair-accumulation**: serious or too frequent damage triggers Misrepair in emergency to maintain the structural integrity and protect from immediate death. However the persistent Misrepair will accumulate and lead to aging.

A structure (molecule, cell or tissue) containing Misrepair has less repair capacity as well as other reduced functions, therefore a misrepaired structure is more fragile to damage, and the appearance of further Misrepair is more likely. So the accumulation of Misrepair is a self-accelerating process. This is the reason why aging phenotypes are not homogenous, which we can understand from the aging dots on the face.

**Aging of tissue**

Aging can occur at different levels: molecular, cellular and tissue. However, an interesting point is whether the aging of tissue, which is directly related to organ function and body lifespan, should be always due to the aging of cells or molecules. The intact function of tissue not only depends on the intact functions of cells/ECMs, but also depends on the intact collaborations (spatial relationships) among them. If a defect from damage or Misrepair) occurs to the tissue structure, which affects the spatial relationships between cells/ECMs, the defect could cause directly a decline of tissue function. If, however, the defect occurs to the cells/ECMs, there can be different outcomes depending upon the location. In tissue that can regenerate, the defected cells/ECMs can be removed and replaced with new cells/ECMs, and this might not necessarily affect tissue function. In tissue that cannot regenerate, such as nerve tissue, the defective cells not only affect the cell-contributed tissue function, but also interrupt the spatial relationships amongst other functional cells/ECMs. Therefore, the residual non-functional cells/ECMs can be also regarded as a defect of tissue structure. For instance, in Alzheimer's disease, it has been realised that the corticocortical disconnection due to the loss of pyramidal neurons, which leads to the communicating failure between different parts of brain is the pathophysiological mechanism of Alzheimer's disease (Delatour, 2004).

The change of tissue structure - the change of the spatial relationship between cells/ECMs, is sufficient for the reduction of tissue function, without requiring the change of the cells or molecules themselves. Therefore aging of body takes place necessarily on tissue level. It is well-known that the common feature of body aging is the deleterious change of tissue structure. Aging does not start only when we are very "old", but it is a lifelong gradual process. One reason why some aged cells can be found in old regenerable tissue is that, the aged cells cannot be replaced in time, or completely, due to the reduced repair function of aged tissue. In this case, the aged cells are the effect of aging of tissue, but not the cause. In this aspect our view is the same as that in the *Multicellular Being Chaos Theory*, which suggests that it is the failure of information transmission in multi-cellular beings between each part that leads to aging (Mulá, 2004), however, this theory does not explain how the breakdown of information transmission occurs.

**The novel theory and its consistence with the physical view on complex system**

*We propose that aging is the result of accumulation of Misrepair of tissue*. Distinguishing between damage and Misrepair clarifies what is the cause (damage) and what is the effect (Misrepair) in the aging process. Misrepair as an imperfect repair links the functions of gene-controlling and damage-exposure in aging. The concept of tissue Misrepair couples aging and development, which obey the same mechanism: tissue-(re)construction. Therefore this theory unifies the multi-factors in aging mechanism.

Likewise, we also predict that the aging of a cell is the result of accumulation of Misrepair of intracellular structure. Many of degeneration forms of cells, such as fatty change, Mallory body, and pigment particle, are kinds of Misrepair, in which the dead components are not degraded, but isolated to prevent their toxicity to cells. The destiny of an aged cell can be, or cleared and replaced by new cell, or isolated by fabric structure when un-degradable, or to transform later into a tumour cell if a critical mutation on DNA occurs.

Our interpretation is consistent with the physical view on complex system. In terms of the Second Law of Thermodynamics, entropy always increases in a closed system. Living system overcomes the entropy increase by metabolism and exchanging energy/substance with environment. Entropy actually characterizes the degree of disorganization. Living beings continuously need to release entropy to protect them from running into thermodynamic equilibrium (death) (Niedermueller, 1990, Haken, 1990, and Wunderlin, 1992). However the entropy increases unavoidably due to the un-removable Misrepair. Therefore it is the Misrepair that accounts for the permanent increase of entropy. In addition, in the spirit of the physical theory of complex systems (Haken, 1990, and Wunderlin, 1992), living beings must be conceived as complex self-organised systems (Kiss, 2009). The characteristics of complex systems are their 'emergent' properties and functions, which are qualitatively different from those of their subsystems. Hence some properties of living beings such as longevity and aging cannot be only explained by the properties of their subsystems such as cells and molecules (Richter, 2002).

**INTERPRETATION OF AGING PHENOTYPES AND LONGEVITY IN THE LIGHT OF THE NOVEL THEORY**

**Aging-related disorders**

**Fibrosis** is a typical phenotype of aging, existing in many old tissues, such as lung, heart muscle and arterial wall, which is unexplainable by the damage-accumulation theory. However, in our view, fibrosis is the result of accumulation of over-produced ECMs or intracellular fibre-like components for repairing, essential for survival by limiting the damage, linking the break, isolating the un-degradable cells or by-products, and reconstructing the structure, which is more robust. For example, in Alzheimer's disease, the neurofibrillary tangles in brain are kinds of fibrosis structures to isolate the damaged components within nerve cells and the amyloid plaques are kinds of fibrosis-like structures with amyloid to isolate the damaged nerve fibres.

**Atherosclerosis** is an aging-related chronic inflammation disease, and the formation of plaques is the result of the over-proliferation and accumulation of macrophages for clearing the lipid in the arterial wall and repairing the endothelium when the endothelium is frequently damaged due to the movements of arterial wall or blood flows. This could be one underlying mechanism of the effect of the caloric restriction on retarding aging (Masoro, 2005). Some studies showed that long-term low-dose of Aspirin is helpful to reduce heart attack (Hung, 2003 and Elwood, 2006). The effect of Aspirin on inhibiting the Atherosclerosis might be one mechanism.

**Aging-related tumours** are the results of mutations due to the accumulation of Misrepair of DNA, which leads to the out-of-control of cell proliferation and finally to tissue disorganization. Actually in many cases, the mutations of DNA cause cell death, and not affect the tissue function. However, the unrepaired or misrepaired DNA damage has been implicated as a causal factor in cancer and aging (Suh, 2006). It is now accepted that the misrepaired DSBs are the main lesions at the origin of both chromosomal abnormalities and gene mutations (Natarajan, 1993 and Bishay; 2001). Misrepair of DNA is one survival mechanism of cell under radiation (Rothkamm, 2002). The lack of cancer in simpler organisms (Campisi, 2000) might be because their lifespans are too short to accumulate sufficient Misrepairs of DNA.

**Wrinkle formation**

It is well-known that the elastic structures are gradually replaced by collagen structures with aging in many organs, such as airway wall, stomach wall, blood vessel wall and skin. On skin this change leads to the formation of wrinkle. Figure 2 shows a schematic representation of the changes of elastic tissue in the process of wrinkle formation. Firstly when the elastic fibres in skin are repeatedly broken, some of them are replaced by collagen fibres by Misrepair; further when the collagen fibres are repeatedly broken during shrinking, some of them are replaced by shorter fibres (Misrepair); with time, more and more elastic fibres are replaced by collagen fibres with different lengths (Misrepair-accumulation), which gradually leads to the reduction of extension potential and the permanent folding of some longer fibres- wrinkle formation (aging).

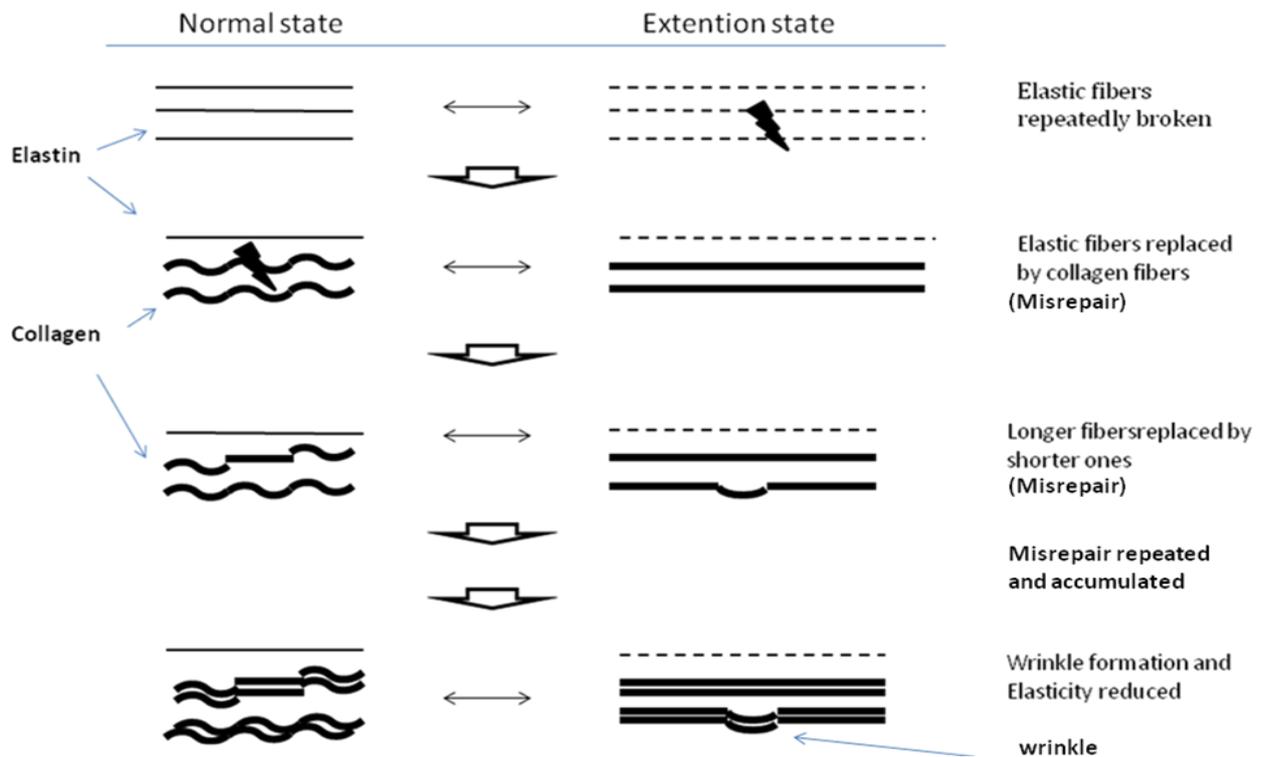

**Figure 2. The formation of wrinkle**

Firstly when the elastic fibres are repeatedly broken, some of them are replaced by collagen fibres by Misrepair. Further some of the collagen fibres are replaced by shorter ones (Misrepair) when the brokens take place during shrinking. With time more and more elastic fibres are replaced by collagen fibres with different lengths (Misrepair-accumulation), which lead to the reduction of extension potential and the permanent folding of some longer fibres - wrinkle.

**Premature (accelerated) aging**

Development and repair are both tissue/body structure formation processes. Any element that interrupts the construction process, e.g. cell division and ECMs production/formation, will affect the development/repair and lead to *mis-construction* (imperfect construction) and *mis-reconstruction* (Misrepair) of tissue/body. At this point, the aging and the development procedures are regulated by a similar genetic mechanism (Zwaan, 2003 and De Magalhaes, 2005). This *mis-construction* and *mis-reconstruction* could be the mechanism of premature (accelerated) aging due to gene mutation. In the pre-matured body, the imperfect tissue structure is with lower function and lower potential, and it goes to failure faster. In this case, it is the accumulation of *mis-construction* (abnormal development) and accumulation of Misrepair together that lead to the failure. Most such individuals cannot survive until full development. The lower repair function, due to gene mutation and abnormal tissue structure, causes more frequent Misrepair. For example, the **Hutchinson-Gilford Progeria** is caused by a gene mutation in the nuclear envelope protein Lamin A (Dechat, 2007). This change of lamin A results in the nuclear blebbing, which interrupts the DNA/RNA synthesis and affects the cell division. In **Fibulin-5-/- aging,** mice with mutation of Fibulin-5 gene have loose and wrinkled skin, vascular abnormalities, and emphysema, all of which are thought to be due to their disorganized and fragmented elastic fibres, which interrupting the development and repair (Hirai, 2007)**.** In **Premature klotho-/- aging,** Kloto gene mutation in mice causes multiple accelerated aging changes due to the disorder of calcium homeostasis, since Klotho gene encodes a type I membrane protein with homology to beta-glycosidase, which is important in the regulation of calcium homeostasis (Lanske, 2007).

**Genes restrict the maximum lifespan by shaping the body**

It is unavoidable that tissue structure goes finally to breakdown due to damage and Misrepair; therefore, the potential of tissue is the key point for longevity. Death often happens from the failure of a key tissue/organ, e.g. the blockage of blood vessel in brain and heart due to Atherosclerosis. The maximum potential of tissue lies in the complexity and maintenance of the tissue structure. More complex structures often have more compensational potential in function due to their repeated and network structures. This could be the reason why species with bigger body/brain size often, although not always, has longer longevity (Aziz, 2005). It will be interesting to compare the structural complexities between different species' to explain the difference of longevity. Genes predetermine this potential by controlling the development. Therefore gene configuration restricts the maximum lifespan by shaping the body. Longer living species' often have a longer development time for their more complex functionality and need longer time for accumulating sufficient Misrepair. This may be the mechanism explaining why different species' have different maximum life-spans. An exception is the big difference in the lifespan between the queen and the worker ants (also in the case of bees): in spite of the same gene configurations. However, the queen and the workers undergo different developments, which lead to different body structures. In *C. elegans*, an alternative (*dauer*) developmental pathway results in a significant longer lifespan (Klass, 1976). These examples indicate that it could be the difference of body structure that finally and directly determines the difference of maximum longevity. Some plants such as most trees, and animals such as ant queen, continue to develop after reproduction, therefore they obtain additional structure complexity and additional function. This increase of complexity enables them to live longer.

It is possible that nature develops longer-living species by increasing the body complexity; however, if the lifespan is too long, individuals would die before the reproduction age, which would lead to the extinction of the species. For example, the Ginkgo biloba tree is reported to be able to live more than 3000 years, and it can survive in extreme weather and environment; however this species is an endangered plant. The survival of species is the result of limited longevity and it is the best evidence of the essentiality of limited longevity.

**REDUCING DAMAGE-EXPOSURE IS THE FEASIBLE WAY TO DELAY AGING-RELATED DISEASES**

Although gene configuration determines the maximum longevity of a species, the individual lifespan is more related to the environmental damage-exposure, and it can only be extended to a certain degree. Damage triggers cell reaction and inflammation, and the repeated damage-caused chronic inflammation is important source of Misrepair. Our theory suggests that for extending lifespan all efforts need to focus on the reduction of Misrepair. This can be achieved by avoiding damage exposure as far as possible, by improving living conditions and medical care. It is especially important to prevent repeated damage and chronic inflammation. To have a regular daily life including healthy diet, sufficient rest and sleep, is important for good reparation. At the same time, we should positively understand aging/Misrepair; without aging/Misrepair, we would have already died from once damage in young age.

An 'improved' repair system by gene-manipulation would theoretically reduce the rate of Misrepair. However the repair-related genes are numerous and they need to act collectively to get their function. Up- or down-regulating of one or two such genes would be insufficient for the goal to improve repairing. In this aspect, the living beings themselves could know better through evolution 'what is the best'. Manipulating these genes without completely understanding their functions could cause fatal danger from side effects. Those genes involved in repairing are also important in development. The extending of lifespan by Knock-out or Knock-in of some genes in animal model such as yeast and *C. elegans*, is probably through redirecting the development (Wei, 2008), and this alteration can potentially cause fatal defects of development (hypoplasia).

Anti-inflammation could be a useful intervention to judge whether Misrepair is the source of aging. From our view, anti-inflammation would reduce the chance of Misrepair, although it would leave some damage to be repaired in delay, which is dangerous. Therefore theoretically the long-term application of low-dose of anti-inflammation medicine would be helpful to reduce aging-related diseases. Some investigations have shown that Aspirin, an anti-inflammation medicine, has effect of 'anti-aging', including cardio-protection and reduction of cancer (Elwood, 2006). In addition, Bulckaen et al have observed the effect of aspirin on anti-aging on animal model (Bulckaen 2008). However the effect is not yet fully confirmed (Kang, 2007). More experiments on animal models are required. One difficulty of such research is the disturbance of the side effects of anti-inflammation to the research result. It has to be emphasized that the side effects of anti-inflammation could cause fatal problems such as bleeding, infection, and organ failure due to insufficient defense and repair. Therefore the medicine, even if it is effective, should be only recommended to some older people, at least more than 50 years old, who have high risk of diseases such as vascular diseases or cancer.

**CONCLUSION**

In this paper we have proposed for the first time the link of Misrepair with aging. Aging is the result of accumulation of Misrepair, and aging of body is necessarily on tissue level. Misrepair and aging is active and necessary for immediate survival and species survival. This theory helps to answer the questions, such as why and how we age (for immediate survival and species survival by Misrepair), why aging mechanism has advantage in evolution, and why we have limited longevity. The following predictions can be made from this theory:

- Aging of cell is the result of accumulation of Misrepair of intracellular structure.
- The longer living species' have more complex body/tissue structures, which need more time to develop.
- Inhibiting chronic inflammation would help reduce aging-related diseases, but it would be accompanied with high risk of other diseases.